\documentclass[a4paper,fleqn,usenatbib,useAMS]{mnras}

\usepackage{graphicx}	
\usepackage{amsmath}	
\usepackage{amssymb}	
\usepackage{multicol}        
\usepackage{bm}		
\usepackage{pdflscape}	
\usepackage{journals}
\usepackage{epstopdf}

\def\HI{H~{\sc i}\, }
\def\HII{H~{\sc ii}\, }

\def\kms{$\textrm{km~s$^{-1}$}$}
\def\nb{\textsc{nbursts}}

\usepackage[T1]{fontenc}
\usepackage{ae,aecompl}

\usepackage{newtxtext,newtxmath}

\title[Young stellar clumps in Arp305]{Young stellar clumps in the interacting systems: Arp305 }
\author[A. Zasov et al.]{Anatoly V. Zasov$^{1,2}$\thanks{E-mail:
zasov@sai.msu.ru}, Anna S. Saburova$^1$, Oleg V. Egorov$^1$, Viktor L. Afanasiev$^3$ 
\\
$^1$ Sternberg Astronomical Institute, Moscow M.V. Lomonosov State University, Universitetskij pr., 13,  Moscow, 119234, Russia\\
$^2$ Faculty of Physics, Moscow M.V. Lomonosov State University, Leninskie gory 1,  Moscow, 119991, Russia \\
$^3$ Special Astrophysical Observatory, Russian Academy of Sciences, Nizhniy Arkhyz, Karachai-Cherkessian Republic 357147, Russia \\
}

\begin{document}
\label{firstpage}
\pagerange{\pageref{firstpage}--\pageref{lastpage}} \pubyear{2016}
\maketitle

\begin{abstract}
We present the results of optical long-slit spectral observations of interacting system Arp305 carried out at the 6m telescope BTA at the SAO RAS: the radial variation of gas kinematics and oxygen abundance. This study continues the series of spectral observations of tidal
debris in the interacting galaxies. Here we pay special attention to the star-forming region between the interacting galaxies -- a tidal dwarf galaxy (TDG) candidate. This star-forming system appears to be gravitationally bound or close to this condition. We show that TDG is metal poor in comparison to the parental galaxy.  It can indicate that either the origin of star-forming gas of TDG was very far from the centre of parent galaxy or its gas has been diluted by the accretion of metal poor intergalactic gas. Nevertheless the region of the brightest emission clump in TDG where a current star formation takes place reveals a noticeable colour excess evidencing a local gas concentration. We pay attention that TDG is situated in the region of intersection of two gas flows and is in a process of
formation probably caused by the collision of these  
flows. 
Low velocity difference between TDG, NGC 4017 and the gaseous  bridge between them
evidences that the lifetime of TDG is restricted by the time of  its fall back
onto the parent galaxy.

\end{abstract}


\section{Introduction}
Arp305 is a small group of galaxies dominated by the wide pair of  interacting spiral galaxies of moderate luminosity: NGC 4016 and NGC 4017 having  very close systemic velocities (about 3500 km~s$^{-1}$).  For the adopted distance 50 Mpc the projected separation of galaxies is 86 kpc. A comparison of the optical morphology with the interaction models (\citealt{Hancock2009};  \citealt{Sengupta2017}) led the authors to conclusion that the interaction between two galaxies took place within the last $4\times 10^8$ yr, although the preceding convergences could also take place. 

\begin{figure}

\includegraphics[width=\linewidth]{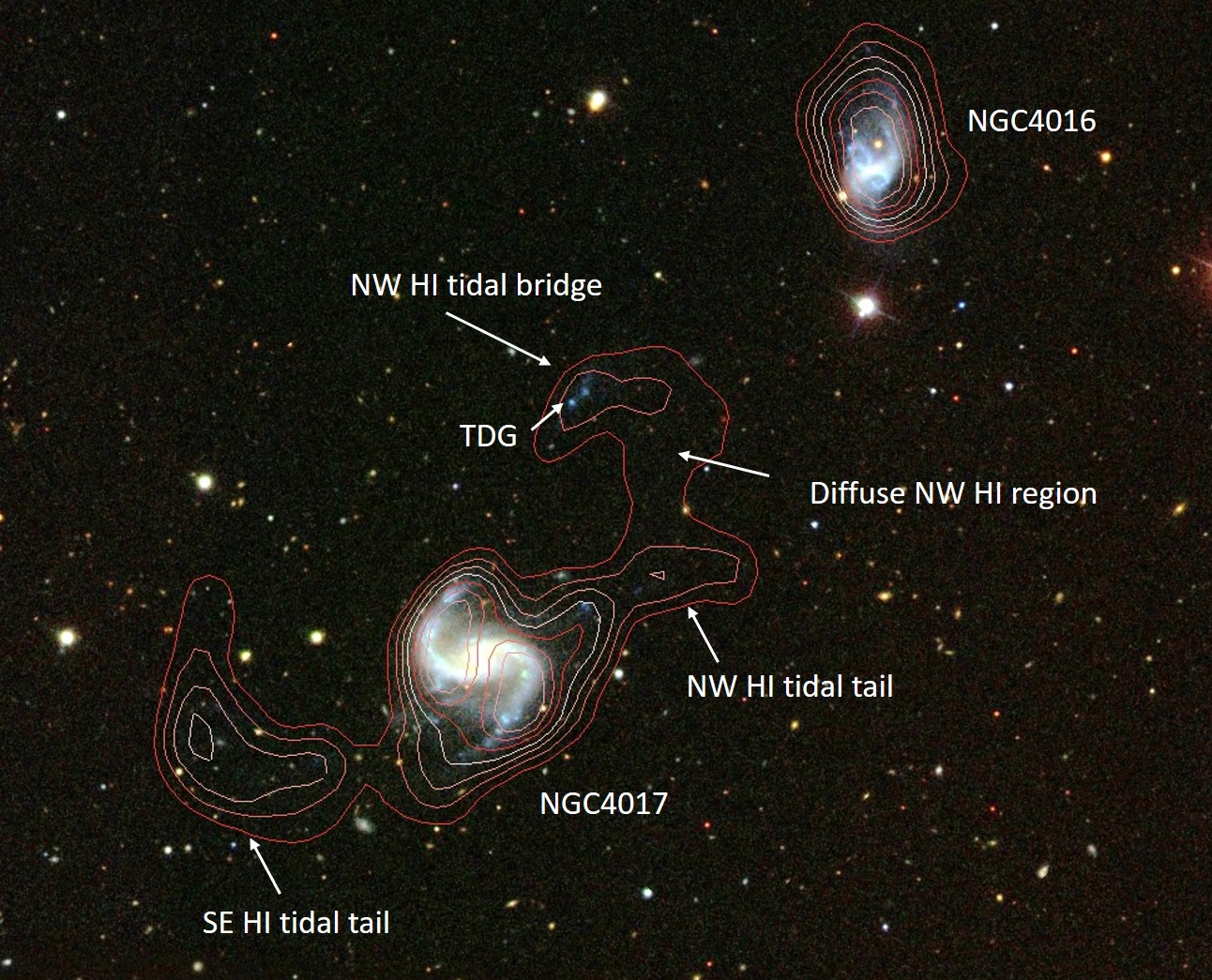}
\caption{\HI surface density contours of Arp305 taken from the GMRT map (\citealt{Sengupta2017}) overlaid on a composite SDSS {\it gri}  image.}
\label{HI}

\end{figure}

Radio observations of neutral hydrogen (H~{\sc i})  which were carried out at WSRT  (\citealt{vanMoorsel1983}) and (with better resolution) at GMRT (\citealt{Sengupta2017}) reveal a vast extension of \HI disc of NGC4017. Indeed, the distribution of \HI in the vicinity of NGC 4017 has a complex character, revealing several tidal features, which evidence that the galaxy is strongly disturbed by the recent convergence with NGC 4016 (\citealt{Sengupta2017}). Higher resolution of GMRT demonstrated the presence of NW and SE \HI tidal tails, diffuse NW gas extension and the bridge between galaxies (or a straight and long tail of \HI directed toward NGC 4016) (see Fig \ref{HI}). 
The largest concentration of \HI in the bridge beyond the optical disc of  NGC 4017  coincides with the optical chain of blue knots immersed in a faint haze, clearly visible in UV, located about halfway between galaxies and stretched  along the line connecting the interacting galaxies at least at  7 kpc (\citealt{Sengupta2017} give 11 kpc for its length). This collection of young stellar objects is considered as  the tidal dwarf galaxy candidate (hereafter TDG). There is also a hint of a continuation of faint optical emission in both directions along the line connecting spiral galaxies.  TDG is especially well-prominent on the UV GALEX maps, being bluer by index (NUV-g), than the parent galaxy NGC 4017 (\citealt{Hancock2009};   \citealt{Smith2010}). Its total $g$ -magnitude is 17.3 (Hancock et al 2009 gave 17.8, the difference is most probably due to uncertainty of the adopted boundaries of TDG) which corresponds to $M_g\approx -16.2^m$.

A photometry of clumps of TDG on the basis of GALEX and SDSS photometry was carried out by \cite{Hancock2009}. They applied the STARBURST99 \citep{Leitherer1999} Simple Stellar Population   models to the photometry data and concluded that the mass of individual young stellar clumps is about $10^6 M_\odot$ by the order of magnitude. The clumps have a very small age, which allows to conclude that the observed stellar island was recently formed from the local concentration of gas in the tidal tail. The existence of older stellar population is the open question. Spitzer infrared fluxes  at 3.6 $\mu $m and 4.5 $\mu$m correspond to the mass of stellar population $~4\times10^7M_\odot$  (\citealt{Sengupta2017}). However this estimate is not reliable, in particular due to the unknown contribution of red giants and warm dust, but even if it is correct, the  mass of gas enclosed in the region of TDG far exceeds the stellar one.

This paper continues the series of spectral observations of tidal debris in the interacting galaxies (Arp 270,  Arp 194,   NGC 4656, see \citealt{zasovetal2015, zasovetal2016, zasovetal2017}). Here  we present the results of optical long-slit spectral observations of NGC4017 and TDG carried out at the 6m telescope BTA at the SAO RAS observatory. The observations were aimed to study dynamics, gas excitation and oxygen abundance in details to try to clarify a possible nature of TDG.

\section{Observations and Data reduction}\label{Obs}
\begin{figure}
\includegraphics[width=\linewidth]{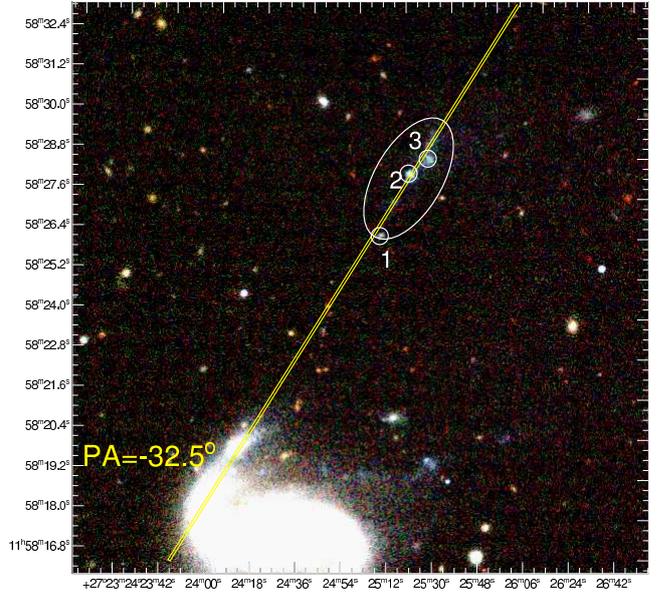}
\caption{The position of the slit on the composite ugr-image. The elliptical contour encircles the position of the proposed TDG between the galaxies. The brightest knots of star formation in this region are shown by circles 1-3. }

\label{map}

\end{figure}

Arp305 was observed in the long-slit mode with  universal spectrograph SCORPIO-2 (\citealt{AfanasievMoiseev2011}) mounted at the prime focus of the 6-m Russian telescope BTA at Special Astrophysical Observatory of the Russian Academy of Sciences. The observations were conducted on 9-th of March of 2016  with the grism VPHG1200@540 which covers the spectral range 3600-7070 \AA ~and has a dispersion of 0.87 \AA ~pixel$^{-1}$.  The exposure time was 1200 sec and the seeing was 1.1 arcsec. The slit orientation is $PA=327.5$. We show the position of the slit on composite {\it ugr}-image in Fig. \ref{map}.

We performed the data reduction in a similar way as in the previous papers \citet{zasovetal2015, zasovetal2016, zasovetal2017}. 
Here we utilized the \textsc{idl} based spectral data reduction pipeline. The data reduction included a bias subtraction, a correction for flat-field and cosmic ray hits removal. The data were calibrated by wavelength using the spectrum of He-Ne-Ar lamp. The uncertainty of the wavelength solution does not exceed  0.05 \AA. We performed the air-glow lines subtraction using the peripheral regions of the slit where the contribution of stellar light of galaxies is negligible. Then we  transformed the peripheral night sky spectrum into the Fourier space and  extrapolated it onto the galaxy position by using the polynomial representation at a given wavelength. As a final step we performed the flux calibration using the spectro-photometrical stellar standard BD33d2642.

We estimated the instrumental profile of spectrograph from the spectrum of twilight sky, taken at the same night. This spectrum was fitted by a broadened high-resolution spectrum of the Sun. The resulting parameters of instrumental profile and their variation both with the wavelength and along the slit were utilized in the spectral fitting by convolving the profile with a grid of stellar population models.

To extract the emission lines from the stellar background  and to obtain the properties of stellar population we firstly binned the spectra using the adaptive binning algorithm in order to achieve minimal signal-to-noise ratio $S/N=15$. After that we used the \nb ~full spectral fitting technique \citep{Chilingarian2007}. This technique enables to fit the observed spectrum in the pixel space against the population model convolved with a parametric line-of-sight velocity distribution. We utilized the high resolution stellar PEGASE.HR \citep{ LeBorgne2004} simple stellar population (SSP) models based on the ELODIE3.1 empirical stellar library. \nb ~technique allowed us to obtain the parameters of stellar population (velocity, velocity dispersion, age and metallicity)  by means of non-linear minimization of the quadratic difference chi-square between the observed and model spectra. After that we subtracted the models of stellar spectra from the observed ones and got the pure emission spectra. We fitted the emission lines by Gaussian distribution and, proceeding from this,  the velocity,  velocity dispersion and emission lines fluxes of the ionized gas were derived.

 \section{Results of Observations}\label{res}

\begin{figure*} 
	\includegraphics[width=\linewidth]{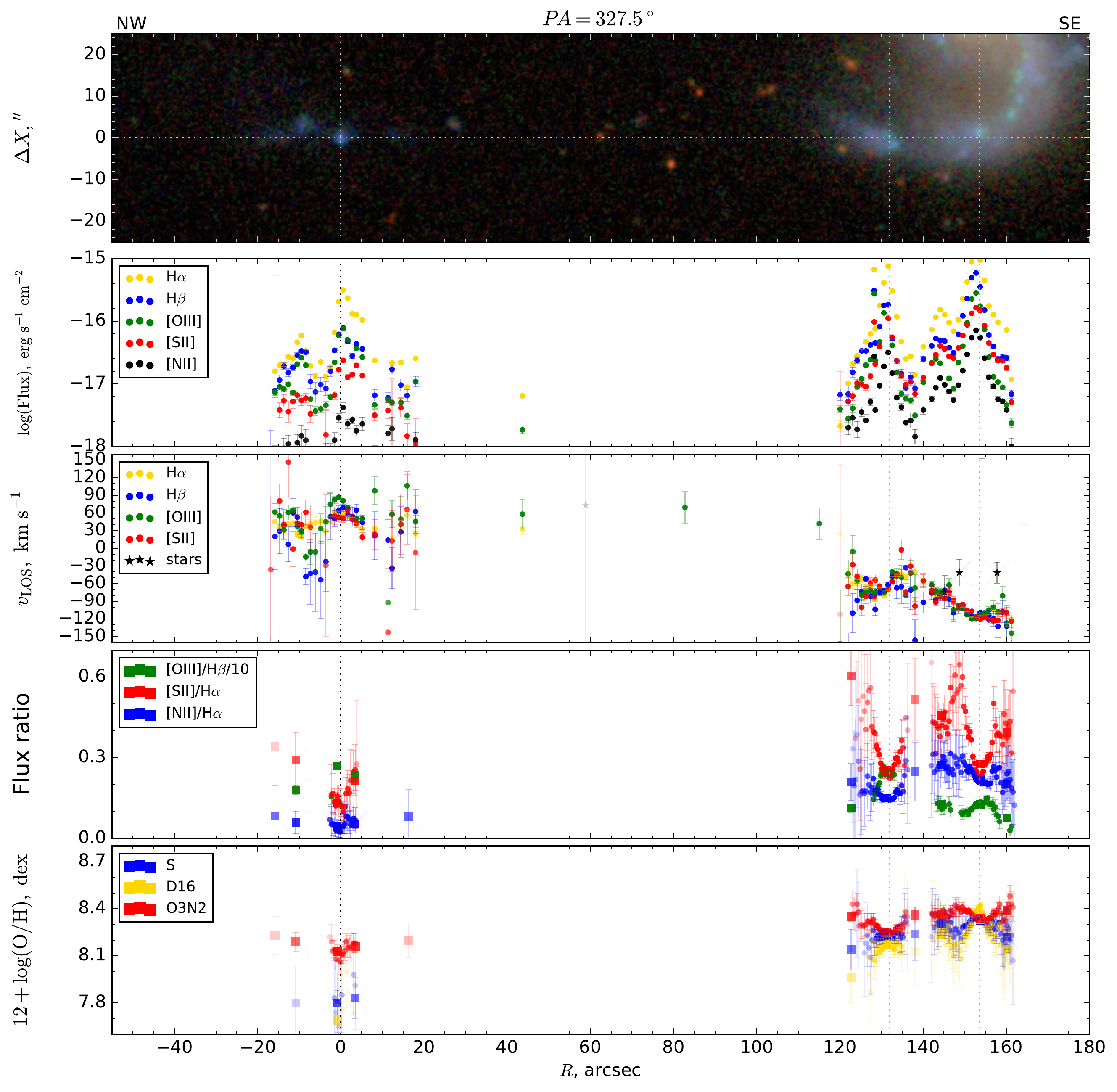}
\caption{The distributions along the slit of: a) emission lines fluxes, b) LOS velocities, c) flux ratios, and d) oxygen abundance. The slit position is shown in the upper panel. Coloured circles correspond to the different emission lines. Asterisks demonstrate the stellar kinematical data. Square symbols show the values obtained after stacking the low brightness emission regions beyond the prominent knots in the scale of  about 10-15''. Pale-colored symbols mark the less certain estimates.
 }
\label{results}
\end{figure*}

The main results of data processing are illustrated in Fig \ref{results}. The panels from top to bottom demonstrate the reference {\it gri} image, radial profiles of the measured fluxes for some bright emission lines, the velocity distribution of gas and stars, the ratios of the emission lines fluxes, the oxygen abundance estimate. Square symbols on the two lower panels show the values obtained after stacking the spectra of extended areas of diffuse emission or of individual \HII clumps (typical scale of about $10-15$ arcsec), while circles correspond to individual pixels after $3\times1$ binning along the slit. Circles and asterisks correspond to the ionized gas and stellar kinematic data.

\subsection{Gas kinematics of the TDG and tidal tail}
The line-of-sight (LOS) velocity distribution of emission gas  reveals a general trend of velocity to grow from NGC4017 to TDG and nearly constant velocity of emission gas between them. A comparison with the velocity field of \HI obtained by \cite{Sengupta2017}, who used two resolution beams of about a quarter and about a half of angular minute, shows a good agreement. High spatial resolution of the optical spectroscopy allowed to reveal local variations of emission gas velocities along the slit caused by non-circular gas motion, which is natural to expect in the regions of the current star formation.  The area of the  significant velocity gradient on the right side of the velocity profile corresponds to the bright region of intense star formation located where the spiral arm of NGC 4017 transforms into the tail. This is the example of so-called `henge clumps' (\citealt{Hancock2009}; \citealt{Struck2011}). Such features within a galaxy disc may  arise near the base of tidal tails, where the crossing zones (caustics) of stellar orbits were  formed  as the result of intersecting stellar  streams.  Gas orbital motion retards there, a  dissipative gas piles up in such regions, and, as a result a  convergence of gas flows triggers the enhanced  star formation ( \citealt{Struck2012}; \citealt{Smith2016}). 

Gas velocities in the TDG region are more disturbed  than beyond it, so that the range of pixel-to-pixel velocity spread exceeds 50 \kms.  At the same time, we observe neither noticeable rotation nor expansion of TDG: the mean velocities of its two brightest clumps differ by no more than 20 \kms. The average weighted velocity dispersion within TDG after taking into account the uncertainties of measurements  is 17 \kms. The restricted spectral resolution does not allow to estimate reliably  velocity dispersion of emission gas from the width of emission lines, but we found  that its upper limit is about 50 \kms, which is quite normal for emission regions with photoionization mechanism of excitation.
\subsection{Gas excitation and oxygen abundance}

\begin{figure}
	\includegraphics[width=\linewidth]{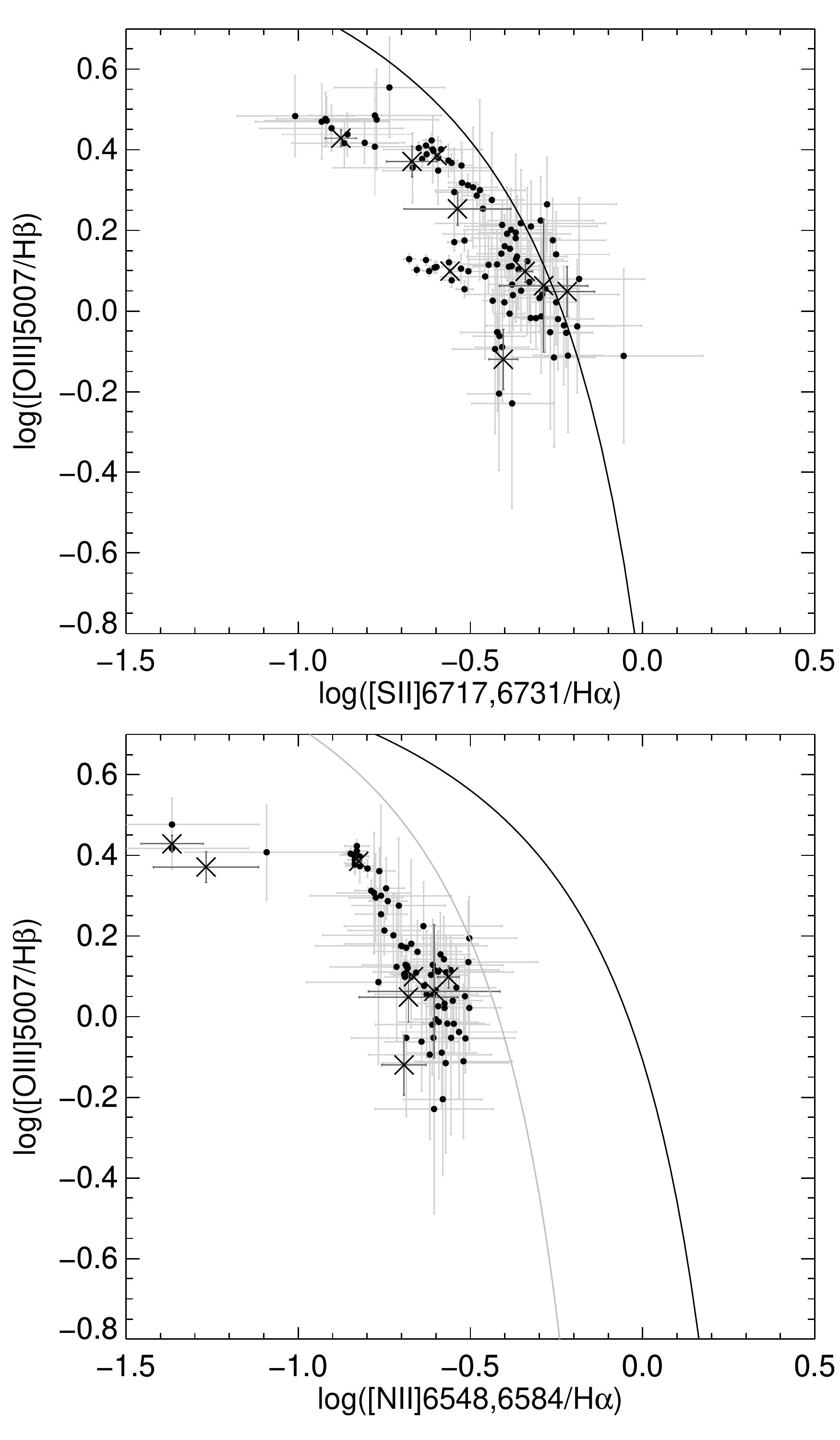}
	\caption{BPT-diagrams constructed for individual bins along the slit (circles) and for stacked spectra of \HII regions or extended area of diffuse gas (crosses; correspond to squares on Fig.~\ref{results}). The `maximum starburst line' \citep{Kewley2001} separating photoionised \HII-regions and all other types of gas excitation is shown in black; grey line from \citet{Kauffmann03} separates the regions with composite mechanism of excitation.}
	\label{fig:BPT}
\end{figure}

  A position of the regions crossed by  the slit  on a classical BPT diagnostic diagram \citep{BPT}  is shown in Fig \ref{fig:BPT}. The demarcation line separates the photoionised \HII regions and the regions which are ionized by a harder radiation (a diffuse ionized gas, see (\citealt{Zhang2017})) or by collisional mechanism (\citealt{Kewley2001}, \citealt{Kauffmann03}). The diagram evidences a photo-ionizing mechanism of \HII-emission in the vast majority of areas. Several points revealing a harder excitation belong to the region of relatively low brightness in the disc of NGC 4017 between the conspicuous  star-forming area (the ``henge clump'') located  on the continuation of spiral arm,  and the main body of the galaxy disc  where most of emission probably belongs to a diffuse ionized gas excitated by hard UV radiation. 

 Line intensity ratios were used to find the oxygen gas-phase abundance, which is the indicator of gas metallicity. The ratios were corrected for selective absorption from the Balmer ratio $H_\alpha / H_\beta$ (see below for more details on the colour excess). Note that the foreground Galactic extinction  for Arp 305 does not exceed $0.1^m$ even in the {\it u} range. 
 
We were not able to use a `direct' $T_e$ method  because of the faintness of the  [O ~\textsc{iii}] 4363 A emission line which is sensitive to electron temperature. Therefore, we used several calibrations that based on the strong emission lines. It is well known that there is a large discrepancy between the metallicity estimates obtained by different methods
\citep[see, e.g.,][]{kewley08, lopez-sanchez12}. Here we selected three methods: S-method proposed by \cite{pilyugin16}, $izi$ method \citep{izi}, and O3N2 method \citep{Marino2013}.
  
 As one can see from the diagram in Fig \ref{results}, the gas abundance in TDG is lower than in the outer part of the main galaxy, crossed by our slit, especially if to accept the S-method. While the value 12+log(O/H) $\sim$ 8.2 - 8.4 for NGC 4017 is compatible with its absolute magnitude  $M_B\approx$ -21, the oxygen abundance of TDG found by izi and O3N3 methods is 8.1 - 8.3, which is too high for its absolute magnitude $M_B\approx -16$ (see the diagrams for dIrr-galaxies presented in  \citealt{Izotov2018} or \citealt{Bergetal2012}, their fig.3,  which predict 7.75 and 8.03 for a given luminosity respectively). Note, that O3N2 method works well only for 12+log(O/H) > 8.15, and hence cannot provide reliable estimates of oxygen abundance in the case of a low metallicity of TDG. On the other hand,  S-method gives 12+log(O/H) $\approx$7.8-8.1, which agrees within the error with the TDG  absolute magnitude. However it should be taken into account that the mass of \HI in TDG exceeds its stellar mass significantly according to \cite{Sengupta2017}, so one would expect a much lower gas enrichment for this dwarf in comparison with the ``normal'' dIrr galaxies with long duration of  star formation, rather it should be closer to metal-poor BDGs (\citealt{Izotov2018}). 
 
\section{On the age of stars in TDG}

\begin{figure}

\includegraphics[width=\linewidth]{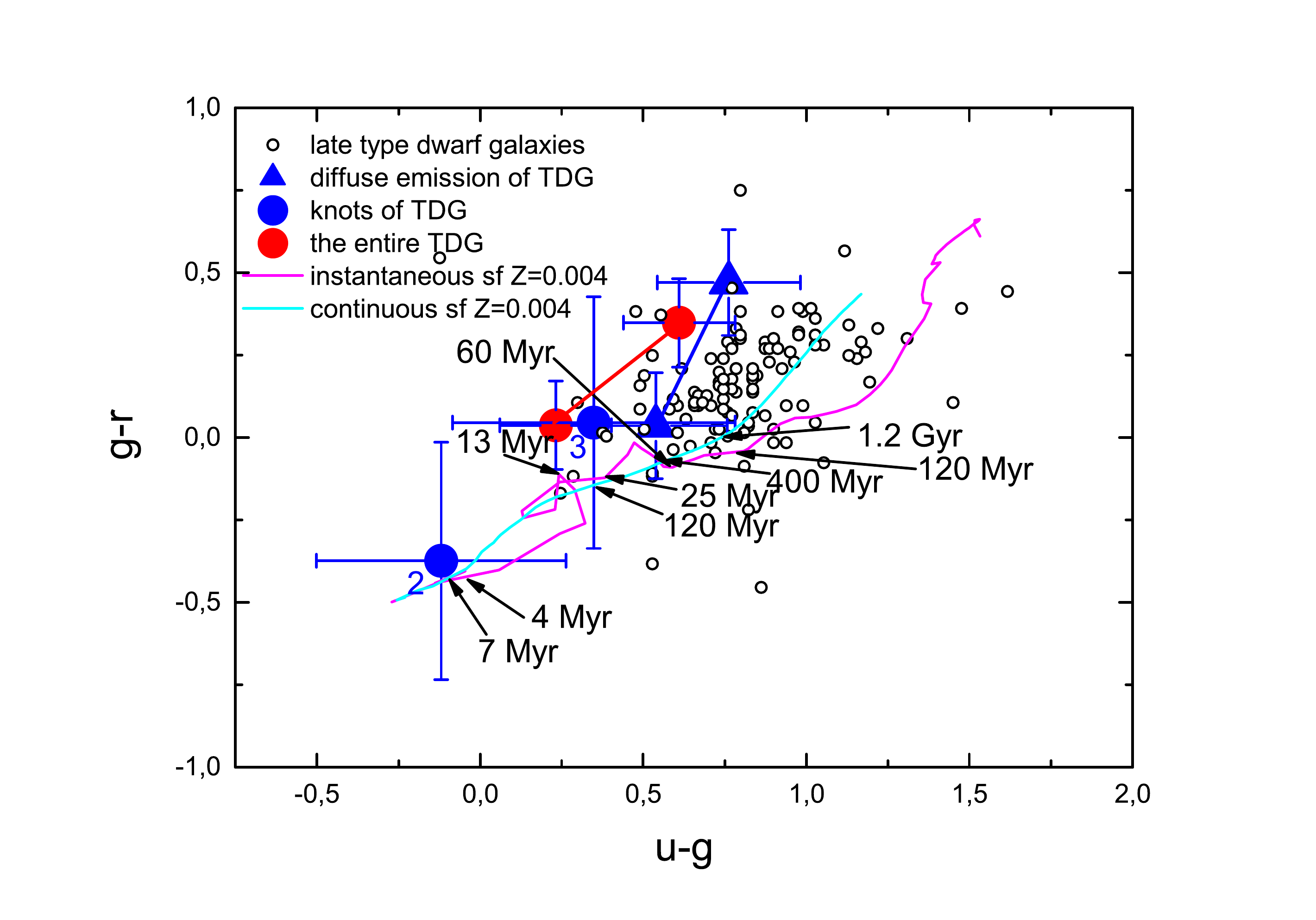}

\caption{The (u-g) vs (g-r) colour diagram. Large symbols with error bars correspond to the regions of star formation between the galaxies (TDG): stellar clumps (2) and (3) (blue circles), the entire island (red circle) and the diffuse emission of the region (the total fluxes of the area with subtracted fluxes of the knots, blue triangle).  Lines give PEGASE2 model tracks with Z=0.004 for instantaneous and continuous star formation (for more details see the text). Small open circles show the position of dwarf irregular  galaxies. The symbols connected by straight lines correspond to the same regions corrected and uncorrected for possible extinction.    }

\label{colordiag}

\end{figure}

In order to specify  the properties of stellar population of TDG we plotted its regions on the (g-r) vs (u-g) diagram (see Fig.\ref{colordiag}). The colors were calculated using SDSS images and corrected for the internal extinction found from the $H_\alpha/ H_\beta$ ratio. The extinction is not zero and has a local maximum: $E_{B-V}= 0.27 \pm 0.05$ in the region of the brightest clump (2) \footnote{See Fig. \ref{map} for the position of the clumps.}, which corresponds to the extinction $A_V = 0.5-1^m$ for usually accepted $R_{B-V}$. This value is not compatible with the observed low column density of gas, especially if to take into account its low metallicity. However it may be explained by high concentration  of gas in the vicinity of young  clusters in a stage of formation at a scale of several hundred pc, which is not resolved by \HI observations. Note that the  colour excess found from $H_\alpha/H_\beta$ of TDG beyond the maximum region of clump (2) is small: $E_{B-V} = 0.03\pm0.05$. The same low extinction is found for emission regions of the main galaxy crossed by the slit.

We considered separately the clumps of star formation 2 and 3\footnote{ Clump 1 was not considered on the diagram because of uncertain estimates of {\it u}-magnitude, however it was taken into account in the photometric estimates of TDG.}  (blue circles), the entire TDG region (red circle) and the diffuse emission of TDG in total resulting after subtraction of  fluxes of the knots 1,2,3 (blue triangle). Since the extinction is inhomogeneous along the slit, which makes the correction uncertain for the entire TDG region, we plot both corrected\footnote{ We used here the maximal value of the correction $E_{B-V}= 0.27$}  and uncorrected colours for the entire TDG and its diffuse emission. In Fig.\ref{colordiag}  we connect these values by the straight lines.  For clump (2) we used $E_{B-V} = 0.27$ and for clump (3) $E_{B-V} = 0.03$.     The two curves demonstrate the PEGASE2 evolutionary
tracks with $Z=0.004$  for continuous star formation (cyan line) and instantaneous star formation (SSP; magenta line) with the Kroupa IMF \cite{Kroupa2002}. Note that the regions of the tracks for ages T<6 Myr are subjects to significant influence of emission lines on the colour indices \citep{smithetal2008}. We also give the ages of stellar population for the closest points of model tracks to the positions of the considered regions.

 For comparison with normal star-forming galaxies we also put the sample of  dwarf irregular galaxies (with $M_B\geq -17^m$) on the diagram, using their UBV colours from Hyperleda database \footnote{http://leda.univ-lyon1.fr/} \citep{Makarov2014}. Their colours in most cases may be explained satisfactory by continuous star formation in the presence of moderate extinction not fully taken into account by Hyperleda, or by the mixture of continuous SF and SSP. As it is expected, Fig.\ref{colordiag} shows that the bright clumps are younger than the entire island. The position of the brightest clumps (2,3) could be satisfactory reproduced by both continuous and instantaneous star formation models. In both models the clump (2) is the youngest one: its age does not exceed 7 Myr. Most probably, this is the star cluster we observe at a stage of formation. Note that if there were no significant extinction, clump (2) would lie close to clump (3) on the diagram. Clump (3) has redder colours and its position is in agreement with the model track for instantaneous star formation for the age of several dozens of  Myr or continuous star formation for roughly hundred of Myr.

 For the TDG in total and for its diffuse emission region beyond the bright young knots the situation is  more complicated. Their colours lie closer to the position of the irregular galaxies and also agree within the uncertainty of the estimates with the instantaneous star formation model track for the age of several dozens Myr, especially if to apply a non-zero correction for colour excess. At the same time, a proximity to the region occupied by dwarf Irr-galaxies evidences that we can not exclude that it may also contain much older stars (with ages of several Gyr).  Note that \cite{Sengupta2017} found that the Spitzer 3.6 $\mu$m and 4.5 $\mu$m flux densities of TDG agree with the presence of old stellar population, formed before the latest interaction of the pair, although the interpretation of [3.6-4.5] colour index remains ambiguous.

Thus, the colours of bright stellar knots generally are compatible with models of star formation which begun  4-25 Myr ago. This age estimate in general agrees with that of \cite{Hancock2009} who obtained age <10 Myr for stellar clumps of TDG,  despite they used different evolutionary tracks and the regions they considered differ from that studied in the current paper (our regions have smaller apertures  of about 4-4.6 arcsec and their positions do not exactly coincide  with respect to that in \citealt{Hancock2009}).  At the same time there are also hints on the presence of much older stars which contribute to the diffuse emission of the stellar island.

\section{Discussion and conclusions}\label{Discussion}

In the case of  Arp305  we observe a current or recent star formation in the extended area of tidal debris between two interacting galaxies, where  
the stellar island rich of gas, or TDG, is the most prominent low brightness object.   

We observe neither noticeable rotation nor expansion of TDG: the mean velocities of its two brightest clumps are very close, however the variation of LOS velocity along the slit is not monotonous inside of TDG. Velocities found from $H\alpha$ line, having the lowest errors, demonstrate that the regular velocity change  within TDG does not exceed 20 \kms, although the rotational velocity may be partially smoothed by the velocity dispersion which is of the same order of magnitude.  The mean-square deviation  of local LOS velocities (in a scale of several pixels) from the error-weighted mean velocity of TDG  after taking into account the mean error of measurements  is $\sigma_r=17 $ \kms. It is essential that the velocity dispersion of \HI has a similar value. 

If to assume that the velocity dispersion is isotropic and the diameter of TDG is $2R\approx 10 $ kpc we obtain the virial mass 
$ M_{dyn} \approx 5\sigma_r^2R/G = 2\times10^9$ M\sun. The mass estimate will be three times lower if the velocity dispersion is mostly oriented along LOS. Even in this non-real case $ M_{dyn}$ remains several times higher than the stellar mass calculated from the r-band luminosity $L_r \times M/L_r = 3.5\times10^8$ M\sun, where $M/L_r=1.2$ M\sun/L\sun ~is stellar mass-to-light ratio obtained from the g-r colour and the model $M/L$-colour relation from \cite{ Belletal2003}. 
However these values of $ M_{dyn}$ are comparable with the observed mass of gas connected with TDG: according to \cite{Sengupta2017} neutral hydrogen mass  $M_{HI}$ = 6.6$\times10^8M_\odot$. After taking into account 30$\%$ input of He the total gas mass: $M_{gas} \approx 10^{9}M_\odot$. It confirms that TDG is at least close to be gravitationally bound, so it is naturally to expect star formation in the most dense regions near the tidal tail axis, where gas pressure is locally enhanced.

As it was shown in previous section, TDG is the low abundant system, where the oxygen  abundance is less (and probably much less) than 1/3 solar value, which is usually taken as  $\mathrm{12+\log(O/H) = 8.69}$ \citep{Asplund09}. To check whether  the metal abundance of TDG can be a result of self-enrichment, it is worth to estimate the effective stellar yield of oxygen $y_{eff}$ needed to account for its metallicity through the process of self-enrichment. For this purpose it is convenient to use the simple expression following from the ``closed box'' evolution model of oxygen enrichment after the formation of stars with the total  mass $M_*$ : 
 \begin{equation}\label{eq1}
y_{eff}=\frac{12\,\left(O/H)\right)}{\ln\left(1/\mu\right)}\,,
\end{equation}
where~$\mu=\frac{M_{\rm{gas}}}{M_{\rm{gas}}+M_*} $,~is the relative  mass of gas, so that 
\begin{equation}\label{eq2}
y_{eff}\approx\frac{12\,\left(O/H\right)}{M_*/1.3M_{HI}}.
\end{equation}.

If we assume for TDG that $M_*\approx 3.5\cdot10^8M_\odot$, (see above), then 
 to account for abundance 12+log(O/H)>7.8, obtained by the bright emission lines methods,  it needs the oxygen yield  $y_{eff}$ >$2\times10^{-3}$. For the higher value 12+log(O/H) = 8.2, which is given by O3N2 and izi methods, $y_{eff} \approx 5\times10^{-3}$. For comparison, the estimates of $y_{eff}$ for the gas-rich galaxies: $y_{eff}\approx (3-6)\times10^{-3}$ (\citealt{Pilyugin2007}; \citealt{Dalcanton2007}; \citealt{Gavilan2013}).  This conclusion agrees with the scenario where the star-forming gas in TDG has chemically evolved independently of the parent galaxy. Note that the outer regions of NGC 4017 are more metal-rich than TDG. One may conclude that either the origin of star-forming gas of TDG was very far from the center of spiral galaxy, so this stellar island is mostly self-enriched, or its gas has been diluted by intergalactic metal-poor gas after stripping. 

  A column gas density in the region of TDG is maximal in the tidal bridge, reaching about $4\times10^{20} g~ cm^{-2}$ as it follows from the \HI surface density map from GMRT. This value is close to the threshold of  appearance of young stellar population in the outer discs and tidal debris at a kpc-scale (\citealt{Maybhate2007}). However the question remains what caused the local burst of star formation we observe in TDG.  The width of star-forming region of TDG perpendicular to the bridge axis in the sky plane is 4- 5 kpc, so we can admit that the extension of gaseous bridge along LOS certainly exceeds 3 kpc. Hence its column density corresponds to a volume density that does not exceed several $10^{-25} g ~ cm^{-3}$,  a density typical for the distant outer regions of stellar discs in gas-rich galaxies. Following the results of \cite{Sengupta2017}, we will take the value 10 \kms ~for the velocity dispersion of gas along LOS in the unperturbed regions of tidal tails, which corresponds to 17 \kms ~for spatial isotropic velocity dispersion.  A rough estimate of Jeans mass following from  these parameters is about $6\times10^8 M_\odot$, which agrees with the total (mostly gaseous) mass of TDG, as it is expected in the case when the mass of TDG is close to that required for gravitational bounding.  Note however that the free-fall time for this density is too long -- about $10^8 yr$, being comparable to the time of dynamic evolution of the perturbed galaxies. Hence, either a gas which fills TDG was inhomogeneous initially, or the current star formation is triggered by some external pressure. 
  
  It is essential that the observed chain of UV clumps and \HII regions in TDG is extended along the bridge and is evidently   formed by gas compression acting perpendicular to the bridge axis. In this connection it is worth noting that TDG is located in the region of visible intersection of two gas flows: a gas forming the bridge and more rarified  gas of diffuse NW \HI region, which is the extension of NW \HI tidal tail (see Fig \ref{HI} in current paper and fig. 4 in \citealt{Sengupta2017}). These two gaseous tidal features possess a different dynamic properties and evidently were formed independently: while the LOS velocity of gas along the  bridge is close to the central velocity of NGC 4016, a  velocity of diffuse gas tail first grows with distance from the parent galaxy, then smoothly falls, approaching to the speed of TDG.  It is natural to propose that these gas flows intersect spatially  in the region where  we observe TDG now. Indeed,  we do not see any sign of continuation of diffuse flow after intersection on the other side of the bridge. In this case  the enhanced gas density and turbulent motions observed in the TDG region as well as the very formation of TDG  may be considered as the recent event  caused by the collision of two gas flows.
  
   A presence of young massive stellar clusters (clumps) in TDG is tightly connected with  the enhanced  velocity dispersion, or turbulent pressure of gas. As it was argued by \cite{Elmegreen1997}, a formation of massive stellar clusters takes place in the regions of enhanced pressure, caused either by high initial density of gas, or turbulent compression and large scale shocks. Parsec-scale simulations of  interacting galaxies also demonstrate that the tidal forces intensify the compressive turbulence leading to the excess of dense gas and generating the star formation activity (\citealt{Renaud2014}), what we observe in the case of blue stellar clumps of TDG. In this case it is hard to expect the presence of stellar substrate around the clumps older than several $10^7$ yr, so the faint diffuse glow of TDG observed in the red/infrared wavelength bands should also be  associated with the young stellar population. Note that  if the assumption of the existence of older stars proposed by \cite{Sengupta2017} is correct,  it would mean that TDG was formed after  the previous  convergence of galaxies and its star formation is recently re-animated. 
   
Our data allow to conclude that most probably   we observe TDG in a process of formation.   Judging from its mass and the observed  spread of velocities of  gas, TDG is close to the virial equilibrium,  although it is not totally virialized yet.  A difference between the velocities of young stellar clumps, found from the brightest emission lines, does not exceed the velocity dispersion of gas, so they are likely to form a single system.  In the proposed scenario it is difficult to expect the long lifetime of TDG. Low velocity difference between
TDG,  NGC 4017 and the gas in the bridge between them evidences that the lifetime of TDG should be short enough, being restricted by the time of its falling on the galaxy -- either radially, along the bridge, or more slowly, if the tangential velocity of TDG, which remains unknown,  is not negligible\footnote{ It is worth noting that the chain of clumps inside of TDG as well as its general shape is elongated along the line connecting two interacting galaxies, which evidences in favor of low tangential velocity of TDG}.  In the latter case the dynamic friction experienced by TDG  in the halo of spiral galaxy may also play role in shortening the  lifetime. Even in this case the time of the fall will be within 1-2 Gyr if it retains its large  mass.
   It is not surprising that the gravitationally bound tidal dwarfs possessing old stellar population  are practically absent in the interacting systems.

The system Arp305 is similar to other interacting systems Arp194 and Arp270 that we studied earlier \citep{zasovetal2015, zasovetal2016}. All these systems reveal the presence of local regions of the intense star formation in the bridge between the interacting galaxies. The fate of the TDG in Arp305 seems similar to  that of the stellar island in Arp194 -- it will be soon accreted by the parent galaxy. TDG is situated in a transitional region between two gaseous systems like the stellar conglomerates in Arp270 which  are the result of compression of colliding gas flows of galaxies in contact.  The ages of stellar population of star-forming islands are young in all three systems we considered. However the variation of oxygen abundance in these systems seems to be different, since in Arp270 the gas is chemically well mixed while for Arp305 and Arp194 the metallicity of stellar island in the bridge is lower than in the interacting galaxies.  In addition, the effective oxygen yield of the star-forming island in Arp270 is significantly lower than expected from the closed box model in contrast with that of the island of Arp305 (see above). It  allows to conclude that the chemical evolution history of the considered interacting systems is different despite of the presence of common feature -- the clumps of intense star formation in the bridge between galaxies.  However there is one general conclusion for all these systems.  Our studies give evidences in favour of short lifetime of stellar islands between interacting galaxies which seems to be the reason of their more rare occurrence in tidal bridges with respect to the tidal tails.

\section*{Acknowledgements}
The authors are grateful to the referee for helpful comments. The authors thank R.Uklein and  D.Oparin for help in observations at BTA. 
The authors are grateful to dr. Chandreyee Sengupta, who kindly provided the \HI density map of the system. Authors thank E. Egorova for the aid on the evolutionary model tracks. The reduction of the spectral data were supported by The Russian Science Foundation
(RSCF) grant No. 17-72-20119. The authors acknowledge the usage of the
HyperLeda database (http://leda.univ-lyon1.fr). 
Funding for the Sloan Digital Sky Survey IV has been provided by the Alfred P. Sloan Foundation, the U.S. Department of Energy Office of Science, and the Participating Institutions. SDSS-IV acknowledges
support and resources from the Center for High-Performance Computing at
the University of Utah. The SDSS web site is www.sdss.org.

SDSS-IV is managed by the Astrophysical Research Consortium for the 
Participating Institutions of the SDSS Collaboration including the 
Brazilian Participation Group, the Carnegie Institution for Science, 
Carnegie Mellon University, the Chilean Participation Group, the French Participation Group, Harvard-Smithsonian Center for Astrophysics, 
Instituto de Astrof\'isica de Canarias, The Johns Hopkins University, 
Kavli Institute for the Physics and Mathematics of the Universe (IPMU) / 
University of Tokyo, Lawrence Berkeley National Laboratory, 
Leibniz Institut f\"ur Astrophysik Potsdam (AIP),  
Max-Planck-Institut f\"ur Astronomie (MPIA Heidelberg), 
Max-Planck-Institut f\"ur Astrophysik (MPA Garching), 
Max-Planck-Institut f\"ur Extraterrestrische Physik (MPE), 
National Astronomical Observatories of China, New Mexico State University, 
New York University, University of Notre Dame, 
Observat\'ario Nacional / MCTI, The Ohio State University, 
Pennsylvania State University, Shanghai Astronomical Observatory, 
United Kingdom Participation Group,
Universidad Nacional Aut\'onoma de M\'exico, University of Arizona, 
University of Colorado Boulder, University of Oxford, University of Portsmouth, 
University of Utah, University of Virginia, University of Washington, University of Wisconsin, 
Vanderbilt University, and Yale University.

\bibliographystyle{mnras}
\bibliography{arp305}

\label{lastpage}

\end{document}